\documentstyle[emulateapj]{article}

\newcommand{\cd}{cm$^{-2}$}

\newcommand{\kms}{km~s$^{-1}$} 
\def\h2{H$_{2}$}

\def\n01{$N_{01}$} 
\def\t01{$T_{01}$} 
\def\tex{$T_{\rm ex}$} 
\def\nd{\nodata}
\newcommand\col[2]{$#1 \times 10^{#2}$}%
\input epsf

\begin{document}

\title{FUSE OBSERVATIONS OF DIFFUSE INTERSTELLAR MOLECULAR HYDROGEN} 
\author{J. MICHAEL SHULL\altaffilmark{1,2}, 
JASON TUMLINSON\altaffilmark{1}, 
EDWARD B. JENKINS\altaffilmark{3}, 
H. WARREN MOOS\altaffilmark{4},
BRIAN L. RACHFORD\altaffilmark{1},
BLAIR D. SAVAGE\altaffilmark{5}, 
KENNETH R. SEMBACH\altaffilmark{4}, 
THEODORE P. SNOW\altaffilmark{1},
GEORGE SONNEBORN\altaffilmark{6}, 
DONALD G. YORK\altaffilmark{7}, 
WILLIAM P. BLAIR\altaffilmark{4},
JAMES C. GREEN\altaffilmark{1},
SCOTT D. FRIEDMAN\altaffilmark{4}
and DAVID J. SAHNOW\altaffilmark{4}
}
\altaffiltext{1}{CASA, Department of Astrophysical and Planetary
   Sciences, University of Colorado, Boulder, CO 80309}
\altaffiltext{2}{Also at JILA, University of Colorado and National
   Institute of Standards and Technology} 
\altaffiltext{3}{Department of Astrophysical Sciences, Princeton University,
   Princeton, NJ 08544} 
\altaffiltext{4}{Department of Physics and Astronomy, Johns Hopkins University,
   Baltimore, MD 21218}
\altaffiltext{5}{Department of Astronomy, University of Wisconsin, Madison, 
   WI 53706} 
\altaffiltext{6}{NASA Goddard Space Flight Center, Greenbelt, MD 20771} 
\altaffiltext{7}{Department of Astronomy and Astrophysics, 
   University of Chicago, Chicago, IL 60637} 

\begin{abstract}

We describe a moderate-resolution FUSE mini-survey of \h2\ in the
Milky Way and Magellanic Clouds, using four hot stars and four AGN as 
background sources. FUSE spectra of nearly every stellar and extragalactic 
source exhibit numerous absorption lines from the \h2\ Lyman and Werner
bands between 912 and 1120 \AA.  One extragalactic sightline
(PKS~2155-304) with low N(H~I) shows no detectable \h2, 
and could be the ``Lockman Hole of molecular gas'', of importance
for QSO absorption-line studies.  
We measure \h2\ column densities in low rotational states ($J$ = 0 and 1) 
to derive rotational and/or kinetic temperatures of diffuse
interstellar gas. The higher-$J$ abundances can constrain models of the
UV radiation fields and gas densities.  In three optically thick
clouds toward extragalactic sources, we find $n_H \approx 30-50$
cm$^{-3}$ and cloud thicknesses $\sim 2-3$ pc.  The rotational 
temperatures for \h2\ at high Galactic latitude, 
$\langle T_{01} \rangle = 107 \pm 17$~K (seven sightlines) and 
$120 \pm 13$~K (three optically thick clouds), are higher than those 
in the {\em Copernicus} sample composed primarily of targets in the disk.  
We find no evidence for great differences in the abundance or state of 
excitation of \h2\ between sight lines in the Galaxy and those in the SMC 
and LMC.  In the future, we will probe the distribution and physical 
parameters of diffuse molecular gas in the disk and halo and in the 
lower-metallicity environs of the LMC and SMC.  

\end{abstract}


\section{INTRODUCTION}
 
As the most abundant molecule in the Universe, molecular hydrogen (\h2)
comprises the bulk of the mass of dense interstellar molecular clouds
and is the main ingredient of star formation (Shull \& Beckwith 1982).  
The first steps of star formation assemble dense interstellar clouds 
from diffuse gas, but the formation mechanism for giant molecular clouds 
from diffuse diffuse gas is still unknown.  Little is known about the 
distribution of diffuse \h2\ in the interstellar medium (ISM) of our 
Galaxy, and major uncertainties remain about its formation, destruction, 
and recycling into dense clouds and stars.  \h2\ also plays 
a central role in our understanding of interstellar chemistry.

Here, we describe early results on \h2\ in the Milky Way and
the Magellanic Clouds obtained with the NASA {\em Far Ultraviolet
Spectroscopic Explorer} (FUSE) satellite.  The FUSE mission and the
capabilities of its spectrograph are described by Moos et al.~(2000)
and Sahnow et al.~(2000).  Over its mission lifetime, FUSE will probe
hundreds of sight lines through the Galactic disk and halo.  
We expect a large fraction of these sight lines to exhibit absorption 
from the Lyman and Werner rotational-vibrational bands of \h2,
some 400 absorption lines between 912 and 1120 \AA, arising from 
rotational levels $J$ = 0~--~7 in the ground vibrational state.  With
FUSE, our team will map the distribution of \h2\ in the diffuse ISM, 
measure its abundance, state of excitation, and rates of formation and
destruction. In \S~2 we summarize our initial observations and describe 
our analysis methods.  In \S~3 we describe results from 8 targets. In 
\S~4 we summarize our results and discuss future directions of 
FUSE \h2\ research.

\section{OBSERVATIONS AND ANALYSIS} 

Our FUSE observations were obtained from 1999 September to 1999
November during the commissioning phase of satellite operations.  A
summary of the observations appears in Table 1. The resolution of the
spectrograph across the band was R $\simeq$ 12,000 for all
observations, based on studies of sharp interstellar lines of \h2, 
Ar~I, and Fe~II.  The data were prepared for analysis by the FUSE data
pipeline as described in Moos et al.~(2000).  Once the
two-dimensional spectra were extracted to one dimension,
a wavelength solution was applied and the oversampled data were
smoothed by a 15-pixel running boxcar to 30 \kms\ resolution.

In typical interstellar conditions, excited \h2\ quickly decays to the 
ground vibrational state of the ground electronic state
(Black \& Dalgarno 1973). 
Observations of \h2\ with the {\em Copernicus} satellite 
occasionally detected rotational lines up to $J = 7$, and we
search for $J \leq 10$.  However, lines above $J = 4$ are difficult to
detect in diffuse interstellar clouds, with our typical $4\sigma$
limiting equivalent width of 30--40 m\AA\ and the corresponding column
density limit of $\rm N(H_2) \simeq 10^{14}$ cm$^{-2}$.  
We use all available \h2\ lines, except where they are 
blended with other interstellar absorption or airglow lines. Typically 
we neglect the Lyman (6-0) band, which is lost in interstellar 
Ly$\beta$, and the Lyman (5-0) band, which is coincident with 
resonance absorption lines \ion{C}{2} $\lambda1036.34$ and
\ion{C}{2}$^{*}~\lambda 1037.02$ and with O~I $\lambda1039.64$ airglow.

We illustrate our data in Figure 1, which displays the FUSE spectrum of
ESO 141-G55, a Seyfert 1 galaxy at Galactic latitude $b = -26.71^{\circ}$.
We detect one \h2\ Galactic component with N(H$_2) =$ \col{2}{19} \cd.
This spectrum is typical of our FUSE data
both in data quality and the abundance of \h2\ lines.

We employ a complement of techniques to analyze \h2\
absorption. The final products are column densities in 
rotational levels, N($J$), from which we can infer the gas density, UV
radiation field, and \h2\ formation and destruction rates.  For high 
column density absorbers with damping
wings in the R(0) and R(1) lines, we fit the line profiles to derive
N(0) and N(1) and use a curve of growth to derive column
densities for $J \geq 2$.  For absorbers with $\rm N(H_{2}) \lesssim
10^{19}$ cm$^{-2}$, we measure equivalent widths of all
\h2\ lines and produce a curve of growth to infer a doppler
$b$-parameter and N($J)$ -- see Fig.\ 2.  
We produce initial estimates of the column
densities and excitation temperatures by fitting all lines
simultaneously with parameters $\rm N(H_2)$, $b$, and $T$, assuming a
Boltzmann distribution at rotational temperature $T_{01}$ for $J=0, 1$ 
and an excitation temperature $T_{\rm ex}$ for $J \geq 2$.  

The {\it Copernicus} \h2\ survey (Spitzer \& Jenkins 1975;
Savage et al.\ 1977) showed that the molecular fraction, 
$f_{\rm H2} = $2N(H$_2$)/[N(H~I) + 2N(H$_2$)], is correlated with
E(B-V) and with total H column density.  They found that $f_{\rm H2}$
undergoes a transition from low values ($\ll 0.01$) to high values
($>0.01$) at E(B-V) $\approx 0.08$. Other {\it Copernicus} studies 
indicate that the populations in $J$ = 0, 1 and $J \geq 2$ 
cannot always be described by a single excitation
temperature (Spitzer, Cochran, \& Hirshfeld 1974).  
\tex\ is generally higher than \t01\ and reflects the
fluorescent pumping of the rotational levels by incident UV 
radiation (Jura 1975a,b). The ratio, N(1)/N(0), is usually set 
by collisional mixing of the ortho- and para-\h2\ states (Shull 
\& Beckwith 1982), but these rotational levels may not yet 
have reached equilibrium in lower density clouds.

\section{RESULTS FOR INDIVIDUAL SIGHT LINES}

\subsection{Magellanic Cloud Targets} 

We report observations of one star in the LMC and three stars in the
SMC. In the cases of AV 232 and HD 5980, we detect \h2\ in both the
Galaxy and the SMC (at +119 \kms\ and +124 \kms; see Table 2).  
Our survey contains two Magellanic Cloud targets with no detected \h2\ 
at LMC or SMC velocities. Sk -67 111 (LMC) and Sk 108 (SMC) show 
Galactic \h2\ absorption near $v_{\rm LSR} \approx 0$ but
have no detected \h2\ at the MC velocities.  Our
limiting equivalent width of 60 m\AA\ ($4\sigma$) in the Sk -67 111 
spectrum sets limits N(0) $\leq$ \col{2.2}{14} \cd\ and
N(1) $\leq$ \col{3.2}{14} \cd\ from the (7-0)~R(0) and R(1) lines. 
For Sk~108, the limit of 30 m\AA\ yields N(0) $\leq$ \col{1.1}{14} \cd\
and N(1) $\leq$ \col{1.6}{14} \cd.  The absence
of \h2\ in these sight lines may be due to low \ion{H}{1} columns, to
intense UV radiation fields that destroy \h2, or to reduced rates
of \h2\ formation on grain surfaces in warm low-metallicity gas.
Abundances for such gas are given by Welty et al.\ (1999).

The SMC stars HD 5980 and AV 232 are separated by 75$''$ on the sky.
Thus, these sight lines may probe the same interstellar clouds in the
Galaxy and SMC. With similar column densities and rotational
temperatures, the Galactic components appear to arise in the same gas.
However, the AV 232 SMC component exhibits a high rotational
temperature \t01\ = $300 \pm 60$ K and a higher excitation temperature
\tex = $520 \pm 90$ K, compared with \t01\ = $98 \pm 13$ K and \tex =
$300 \pm 60$ K for the HD 5980 absorber. The level populations 
likely arise from differing excitation rates in separate clouds,
$\sim20$ pc apart at the SMC.

The only detections of Magellanic Cloud \h2\ absorption prior 
to FUSE were observations with ORFEUS (Richter et al.~1998;
de Boer et al.~1998), at low signal to noise but comparable 
($R \approx 10^4$) resolution.  With a longer mission lifetime 
and higher effective area, FUSE offers significant improvement 
for \h2\ studies. For example, with higher S/N, we do not confirm 
the claimed ORFEUS detections of \h2\ (J = 5, 6, 7) 
at +160 \kms\ toward HD~5980 (Richter et al. 1998).  
Our observations of \h2\ column densities in the 
+120 \kms\ component differ considerably, owing to a better 
curve of growth (Fig.\ 2) with $b = 10$ \kms\ compared to 
$b = 6$ \kms\ with ORFEUS.  
Based on 28 FUSE lines ($J = 0-3$) compared to 8 ORFEUS 
lines, we derive log N(0) = 15.10 (compared to 16.57) and  
log N(1) = 15.30 (compared to 15.90). Within error bars, 
both studies agree on $N(2)$ and $N(3)$.  Our data yield 
$T_{01} = 98 \pm 13$~K for the SMC gas.

\subsection{Extragalactic Targets} 

Our four extragalactic targets at Galactic latitude $|b| \geq 25^{\circ}$ 
probe a large volume of the Galaxy and its halo and allow us to 
measure the abundance and physical properties of molecular gas outside 
the disk for the first time.  The {\it Copernicus} disk result, 
$\langle T_{01} \rangle = 77 \pm 17$ K (Savage et al. 1977), was drawn 
from 61 stars with $|b| \leq 65^\circ$. From seven FUSE sightlines, we 
derive $\langle$\t01$\rangle$ = 107 $\pm$ 17~K.  In the three
optically thick extragalactic sightlines, $\langle$\t01$\rangle$ = 120 $\pm$
13~K.  These results may indicate a higher level of rotational excitation in 
the upper disk and low halo, which could arise from more intense UV pumping 
or photoelectric heating from grains in the infrared cirrus.
However, these interpretations are still speculative, since the
the $J = 0,1$ levels may not be in equilibrium with the kinetic temperature
as a result of UV radiative pumping.

In general, the \h2\ rotational populations can be modeled (Jura
1975a,b) to provide physical diagnostics of \h2-bearing clouds, 
such as $n_H$, $T_{01}$, thermal pressure, and UV radiation field.
We also derive internally consistent values of N(H$_2$)
and $T_{\rm ex}$ for these clouds, which could indicate the presence of 
warm \h2-bearing gas in the lower Galactic halo.  
For the three AGN sightlines with detectable \h2, we find $n_H$
= 50, 50, and 36 cm$^{-3}$ for Mrk~876, ESO~141-G55, and
PG~0804+761 respectively, which yields cloud thicknesses
of 1.9, 2.4, and 3.3 pc. The line of sight to ESO 141-G55 intersects a 
region of enhanced {\it IRAS} 100 $\mu$m emission  (Sembach, Savage, \& 
Hurwitz 1999). The other extragalactic sightlines lie behind high-velocity 
cloud Complex C (Mrk~876) and weak infrared cirrus (PG~0804+761). Our \h2\ 
results suggest that these clouds are also compressed.

\section{SUMMARY AND FUTURE DIRECTIONS} 

The major results of these early FUSE observations are: (1) the 
ubiquity of \h2\ in nearly all sightlines; (2) the generally
warmer rotational temperatures, $T_{01}$ toward
high-latitude gas; and (3) the presence of one low-\h2\ 
extragalactic sightline (PKS~2155-304).   
We have not detected \h2\ in the high velocity clouds (HVCs)
toward Mrk~876 or PKS~2155-304, although we intend to
continue our HVC searches toward other targets.  The HD molecule
was not detected in any diffuse cloud, but
results from a complementary program (Snow et al.\ 2000;
Ferlet et al.\ 2000) detect \h2\ and HD toward the reddened star HD~73882.
Detections of \h2\ in other FUSE sight lines are analyzed by 
Friedman et al.\ (2000), Mallouris et al.\ (2000), and 
Oegerle et al.\ (2000).

The lack of detectable \h2\ toward PKS 2155-304 is quite 
interesting, in contrast to the general presence of \h2\ at high 
Galactic latitudes.  Extragalactic sight lines with low \h2\ 
are valuable for studies of QSO absorption lines
(Shull et al.\ 2000; Sembach et al.\ 2000), since \h2\ 
line blanketing can add confusion to line identifications (Fig.\ 1).
The observed low N(H$_2$) may reflect the intrinsic patchiness of the 
ISM, but can also be influenced by \h2\ formation/destruction 
in regions of high UV radiation fields and low hydrogen column density. 
PKS~2155-304 has the lowest column density, log N(H~I) 
$\approx 20.15$, of any of our targets.  
The Sk~108 sight line shows no \h2\ at SMC
velocities, and the Sk~-67~111 sight line shows no \h2\ at LMC
velocities. Although \h2\ formation may not reach equilibrium 
with UV destruction in these diffuse clouds, the \h2\ formation rate 
could depend on the metallicity through the content of dust grains, 
whose surfaces catalyze \h2\ formation.    
In the Galactic halo, some clouds may be exposed to strong
UV dissociating radiation from OB associations
in the disk.  With more sight lines, we should be able to correlate 
regions lacking \h2\ with regions of low N(H~I)  
and low metallicity (LMC, SMC).
Close sight lines such as those to HD 5980 and AV 232 present an 
opportunity to probe the size of diffuse \h2-bearing interstellar clouds. 

The primary product of future FUSE diffuse \h2\ studies will be a
map of the distribution of diffuse clouds in the disk and halo of the
Milky Way, using a large sample of sight lines 
toward Galactic and extragalactic targets.  Multiple sight lines to the
Magellanic Clouds will permit the study of CO/\h2\ and possibly HD/\h2\ 
in lower-metallicity environs, including high-velocity clouds in
the Galactic halo and Magellanic Stream (Gibson et al.~2000).
Observations of multiply-intersected absorbers and \h2\ associated with 
infrared cirrus and high velocity clouds can constrain the size of 
these diffuse interstellar clouds.  

\acknowledgements 

This work is based on data obtained for the Guaranteed Time Team by the
NASA-CNES-CSA FUSE mission operated by the Johns Hopkins University.
Financial support to U.S. participants has been provided by NASA
contract NAS5-32985. J.M.S. also acknowledges the support of NASA 
astrophysical theory grant NAG5-4063.




\begin{deluxetable}{lclccccc}
\footnotesize
\tablecolumns{4}
\tablenum{1}
\tablewidth{0pt}
\tablecaption{FUSE \h2\ Mini-Survey Target List}
\tablehead{
\colhead{Target}
& \colhead{Channels \tablenotemark{1}}
& \colhead{$t_{\rm exp}$~(N) \tablenotemark{2}}
& \colhead{S/N \tablenotemark{3}} \nl
& \colhead{}
& \colhead{(ksec)}
& \colhead{}
}
\startdata
Mrk 876            &  L1 L2 S1  & 45.9 (10)   & 15 \nl
ESO 141-G55        &  L1 L2 S2  & 35.8 (20)   & 15 \nl
PG 0804+761        &  L1 L2 S1  & 39.6 (12)  & 15 \nl
Sk -67 111 (LMC)   &  L1        & 13.8 (7)   & 15 \nl
AV 232 (SMC)       &  L1        & 10.3 (5)   & 12  \nl
HD 5980 (SMC)      &  L1        &  3.2 (8)   & 15 \nl
Sk 108 (SMC)       &  L1        & 13.3 (8)   & 30 \nl
PKS 2155-304       &  L1 L2 S2  & 37.1 (22)  & 20 \nl
\enddata
\tablenotetext{1}{Detectors LiF (L1 and L2 channels) and
   SiC (S1 and S2 channels) have wavelength ranges described by
   Moos et al.\ (2000).}
\tablenotetext{2}{Exposure time and number (N) of sub-exposures.}
\tablenotetext{3}{Signal-to-noise ratio per smoothed (30 \kms) bin
   between 1040--1050 \AA.}
\end{deluxetable}
\normalsize



\begin{deluxetable}{lccccccccc}
\footnotesize
\tablecolumns{10}
\tablenum{2}
\tablewidth{0pt}
\tablecaption{FUSE \h2\ Mini-Survey Results \tablenotemark{1}}
\tablehead{
  \colhead{Name (Type)}
& \colhead{$l$ ($^{\circ}$)}
& \colhead{$b$ ($^{\circ}$)}
& \colhead{$v_{\rm LSR}$}
& \colhead{N(H$_2$)}
& \colhead{N(H~I)}
& \colhead{$f_{\rm H2}$}
& \colhead{$T_{01}$}
& \colhead{$T_{\rm ex}$}
& \colhead{$b_{\rm dopp}$} \nl
  \colhead{}
& \colhead{}
& \colhead{}
& \colhead{(km s$^{-1}$)}
& \colhead{(cm$^{-2}$)}
& \colhead{(cm$^{-2}$)}
& \colhead{}
& \colhead{(K)}
& \colhead{(K)}
& \colhead{(km s$^{-1}$)}
}
\startdata
\cutinhead{\em Magellanic Cloud Sight Lines}
Sk -67~111 (LMC, O7) & 277.75 & -32.97 & 0    & \col{2.3}{15} & \col{6.2}{20}
  & \col{7.4}{-6} & 122 & 300 & 8 \nl
  & & & +270 & $<$\col{5.5}{14} & \nd & \nd & \nd & \nd & \nd \nl
AV 232 (SMC, O9)  & 302.06 & -44.94 & 0 & \col{1.6}{16} & \col{3.6}{20}
   & \col{8.9}{-5} &  86 & 195 & 5 \nl
   & & & +119 & \col{2.0}{15} & \nd & \nd & 300 & 520 & 7  \nl
HD 5980 (SMC, WR) & 302.07& -44.95& 0 & \col{9.8}{15} & \col{3.6}{20}
   & \col{5.4}{-5} & 98 & 195 & 5 \nl
   & & & +124 & \col{4.2}{15} & \nd & \nd & 98 & 300 & 10 \nl
Sk 108 (SMC, WR) & 301.63& -44.99& 0 & \col{1.1}{16} & \col{3.4}{20}
   & \col{6.5}{-5} & 85  & 180 & 7 \nl
   & & & +120 & $<$\col{2.7}{14} & \nd &  \nd & \nd & \nd & \nd \nl
\cutinhead{\em Extragalactic Sight Lines}
Mrk 876 (Seyf1) & 98.27 & 40.38  & 0 & \col{2.3}{18} & \col{2.7}{20}
   & \col{1.7}{-2} & 135 & 200 & 5 \nl
ESO 141-G55 (Seyf1) & 338.18 & -26.71 & 0 & \col{1.9}{19} & \col{3.5}{20}
   & \col{9.3}{-3} & 113 & 540 & 10  \nl
PG 0804+761 (QSO) & 138.28 & 31.03 & 0 & \col{6.1}{18} & \col{3.5}{20}
   & \col{3.4}{-2} & 113 & 180 & 8  \nl
PKS 2155-304 (BL~Lac) & 17.73 & -52.27 & \nd & $<$\col{4.6}{14} &
    \col{1.4}{20} & $<$\col{6.6}{-6} & \nd & \nd & \nd
\enddata
\tablenotetext{1}{Columns:  Target name (type); Galactic coordinates
   (e at $l$, $b$); \h2\ absorber velocity in LSR (Galactic \h2\ is assumed
   to lie at 0 km~s$^{-1}$ in LSR); column densities of \h2\
   (typically $\pm 0.2$ dex from FUSE) and H~I (21 cm emission, Dickey \&
   Lockman 1990; Lockman, Murphy, \& Sembach, in preparation; Savage \&
   deBoer 1981; Sembach, Savage, \& Hurwitz 1999); molecular fraction,
   $f_{\rm H2}$ = 2N(H$_2$)/[N(H~I) + 2N(H$_2$)];
   rotational temperature $T_{01}$ (typically
   $\pm 15$~K) from $J = 0$ and 1; excitation temperature $T_{\rm ex}$
   (typically $\pm 75$~K) from $J \geq 2$; and doppler parameter
   (typically $\pm 2$ \kms) from \h2\ curve of growth. }
\end{deluxetable}
\normalsize


\begin{figure*}
\centerline{\epsfxsize=0.8\hsize{\epsfbox{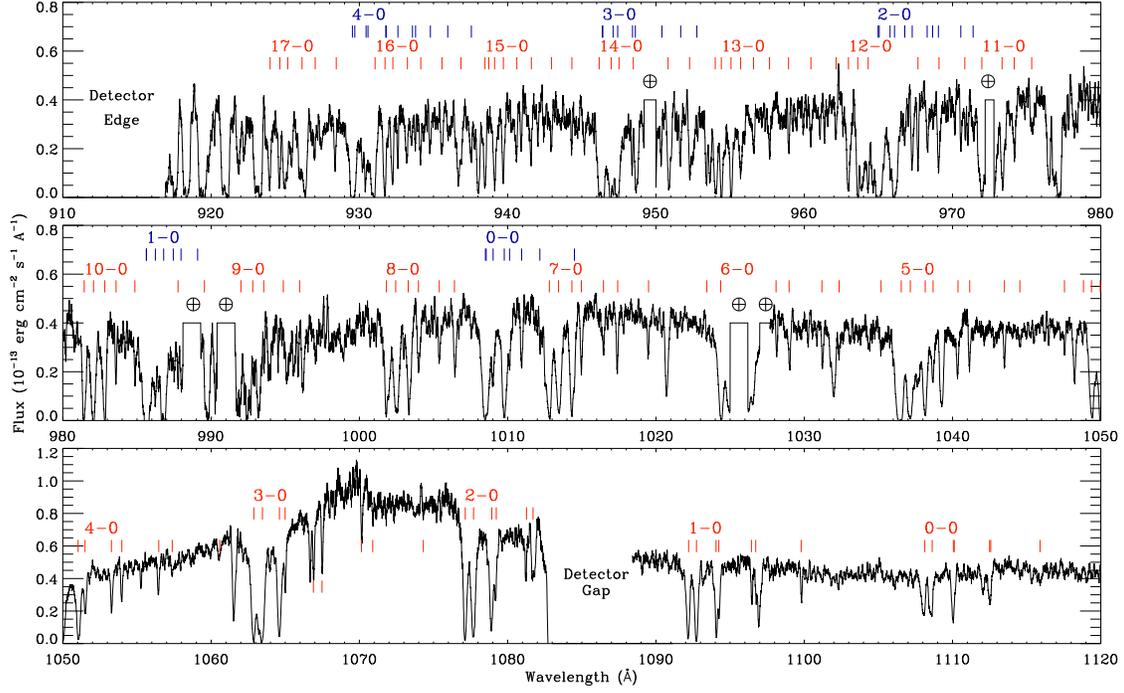}}}
\vspace{0.2in}
\figcaption{FUSE spectrum of ESO 141-G55. Bright airglow lines have
   been truncated.  This spectrum has a resolution of $R \simeq 12,000$
   and $S/N \approx 15$ per smoothed (30 \kms) bin (1040--1050 \AA)
   and $S/N \approx 25$ at 1070 \AA. Lower (red) and upper (blue) ticks
   mark the detected Lyman and Werner lines of \h2, respectively. The
   interstellar and intergalactic lines are analyzed further by Sembach
   et al.~(2000) and Shull et al.~(2000).}
\label{fig2}
\end{figure*}

\begin{figure*}
\centerline{\epsfxsize=0.45\hsize{\epsfbox{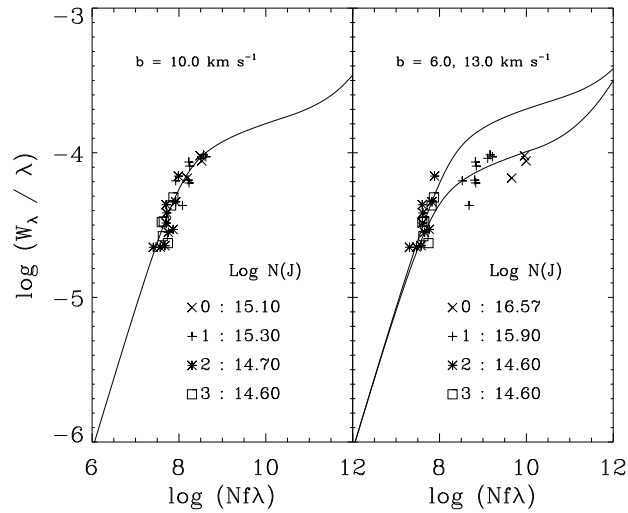}}}
\vspace{0.2in}
\figcaption{(Left) Composite curve of growth for 28 \h2\ lines
   ($J = 0, 1, 2, 3$) in the SMC (+120 \kms) component
   toward HD~5980 with best-fit doppler parameter,
   $b = 10$ \kms, and column densities, N(J). (Right)
   Same 28 FUSE lines but with the higher ORFEUS column
   densities (Richter et al.\ 1998, who used $b = 6$ \kms).}
 \label{fig1}
 \end{figure*}

\end{document}